\documentclass[final]{siamltex}
\usepackage{latexsym}
\newtheorem{fact} [theorem]{Fact}
\newcommand{\Com}{{\mathchoice {\setbox0=\hbox{$\displaystyle\rm
        C$}\hbox{\hbox to0pt{\kern0.4\wd0\vrule height0.9\ht0\hss}\box0}}
{\setbox0=\hbox{$\textstyle\rm C$}\hbox{\hbox
to0pt{\kern0.4\wd0\vrule height0.9\ht0\hss}\box0}}
{\setbox0=\hbox{$\scriptstyle\rm C$}\hbox{\hbox
to0pt{\kern0.4\wd0\vrule height0.9\ht0\hss}\box0}}
{\setbox0=\hbox{$\scriptscriptstyle\rm C$}\hbox{\hbox
to0pt{\kern0.4\wd0\vrule height0.9\ht0\hss}\box0}}}}

\newcommand{\qed}{\hfill$\Box$}
\newcommand{\R}{{\rm I\!R}}
\newcommand{\N}{{\rm I\!N}}
\newcommand{\Q}{{\mathchoice {\setbox0=\hbox{$\displaystyle\rm
Q$}\hbox{\raise 0.15\ht0\hbox to0pt{\kern0.4\wd0\vrule
height0.8\ht0\hss}\box0}} {\setbox0=\hbox{$\textstyle\rm
Q$}\hbox{\raise 0.15\ht0\hbox to0pt{\kern0.4\wd0\vrule
height0.8\ht0\hss}\box0}} {\setbox0=\hbox{$\scriptstyle\rm
Q$}\hbox{\raise 0.15\ht0\hbox to0pt{\kern0.4\wd0\vrule
height0.7\ht0\hss}\box0}} {\setbox0=\hbox{$\scriptscriptstyle\rm
Q$}\hbox{\raise 0.15\ht0\hbox to0pt{\kern0.4\wd0\vrule
height0.7\ht0\hss}\box0}}}}
\newcommand{\Z}{{\mathchoice
{\hbox{$\sf\textstyle Z\kern-0.4em Z$}} {\hbox{$\sf\textstyle
Z\kern-0.4em Z$}} {\hbox{$\sf\scriptstyle Z\kern-0.3em Z$}}
{\hbox{$\sf\scriptscriptstyle Z\kern-0.2em Z$}}}}

\begin{document}

\title{Lower Bounds for Quantum Communication Complexity\thanks{The
  results in this paper have previously appeared in its conference version
  \cite{Kl01} at FOCS'01.}}
\author{Hartmut Klauck\thanks{Address: CWI,
  P.O. Box 94079, 1090 GB Amsterdam, the Netherlands. Email: klauck@cwi.nl.
  Supported by the EU 5th framework
  program QAIP IST-1999-11234 and by NWO grant 612.055.001.}
}

\maketitle

\begin{abstract}
We prove new lower bounds for bounded error quantum communication
complexity. Our methods are based on the Fourier transform of the
considered functions. First we generalize a method for
proving classical communication complexity lower bounds developed by
Raz \cite{R95} to the quantum case. Applying this method we give an
exponential separation between bounded error quantum communication
complexity and nondeterministic quantum communication complexity.
We develop several other lower bound
methods based on the Fourier transform, notably showing that $\sqrt{\bar{s}(f)/\log n}$, for the average
sensitivity $\bar{s}(f)$ of a function $f$, yields a lower
bound on the bounded error quantum communication complexity of
$f(x\wedge y\oplus z)$, where $x$ is a Boolean word held by Alice and
$y,z$ are Boolean words held by Bob.
We then prove the first large lower bounds
on the bounded error quantum communication complexity of functions,
for which a polynomial quantum speedup is possible.
For all the functions we investigate, the only previously
applied general lower bound method based on discrepancy yields
bounds that are $O(\log n)$.
\end{abstract}



\pagestyle{myheadings}
\thispagestyle{plain}
\markboth{H. Klauck}{Lower Bounds for Quantum Communication Complexity}

\section{Introduction}
Quantum mechanical computing and communication has been studied
extensively during the last decade. Communication has to be a
physical process, so an investigation of the properties of physically
allowed communication is desirable, and the fundamental
theory of physics available to us is quantum mechanics.

The theory of communication complexity deals
with the question how efficient communication problems can be solved, and has various
applications to lower bound proofs for other resources (an introduction
to (classical) communication complexity can be found in \cite{KN97}).

In a quantum protocol (as defined in \cite{Y93}) two players
Alice and Bob each receive an input, and have
to compute some function defined on the pair of
inputs cooperatively. To this end they exchange
messages consisting of qubits, until the result
can be produced from some measurement done by one of the players
(for surveys about quantum communication complexity see \cite{T99,B00,Kl00}).

It is well known that quantum communication protocols can sometimes
be substantially more efficient than classical probabilistic protocols:
The most prominent example of such a function is the disjointness
problem $DISJ_n$, in which the players
receive incidence vectors $x,y$ of subsets of
$\{1,\ldots,n\}$, and have to decide whether the
sets are not disjoint: $\bigvee (x_i\wedge y_i)$.
By an application of Grover's search algorithm \cite{G96} to
communication complexity given in \cite{BCW98} an upper bound of
$O(\sqrt{n}\log n)$ holds for the bounded
error quantum communication complexity of
$DISJ_n$. Recently this upper bound has been improved to
$O(\sqrt{n}c^{\log^*n})$ in \cite{HW02}.
The classical bounded error communication complexity of
$DISJ_n$ on the other hand is $\Omega(n)$ \cite{KS92}.
The quantum protocol for $DISJ_n$ yields the largest gap between
quantum and classical communication complexity
known so far for a total function. For partial functions and
so-called sampling problems even exponential gaps
between quantum and classical communication complexity are known,
see \cite{R99, BCW98,ASTVW98}.

Unfortunately so far only few lower bound methods for quantum
communication complexity are known: the logarithm of the rank of the communication
matrix is known as a lower bound for exact
(i.e., errorless) quantum communication \cite{BCW98,BW01}, the
(in applications often weak) discrepancy method can be used to
give lower bounds for protocols with error \cite{K95}.
Another method for protocols with bounded error
requires lower bounds on the minimum rank
of matrices approximating the communication matrix \cite{BW01}. A very
recent result by Razborov \cite{R02} (published subsequently to this
paper) implies such lower bounds for a limited
class of functions, previously such results were unknown. In this
paper we introduce several lower bound methods for bounded error
quantum communication complexity exploiting algebraic properties of
the communication matrix.

Let $IP_n$ denote the inner product modulo 2 function, i.e.,
\[IP_n(x,y)=\bigoplus_{i=1}^n(x_i\wedge y_i).\]

Known results about the discrepancy of the inner product function
under the uniform distribution then imply that
quantum protocols for $IP_n$ with error $1/2-\epsilon$
have complexity $\Omega(n/2-\log(1/\epsilon))$, see \cite{K95} (actually
only a linear lower bound assuming constant error is proved there, but
minor modifications give the stated result). The inner product
function appears to be the only explicit function, for which a large lower
bound on the bounded error quantum communication complexity
has been published prior to this paper.

We prove new lower bounds on the bounded error
quantum communication complexity of several functions. These bounds are
exponentially bigger than the bounds obtainable by the discrepancy method. Note
that we do not consider the model of quantum communication with prior
entanglement here (which is defined in \cite{CB97}).

Our results are as follows. First we generalize a lower bound method
developed by Raz \cite{R99} for classical bounded error protocols to the quantum
case. The lower bound is given in terms of the sum of absolute values
of selected Fourier coefficients of the function. To be able to
generalize this method we have to decompose the quantum protocol into a
``small'' set of weighted monochromatic rectangles, so that the sum of these
approximates the communication matrix. Opposed to the classical case
the weights may be negative, but all weights have absolute value
at most 1.

Applying the method we get a lower bound of $\Omega(n/\log n)$ for the
bounded error quantum communication complexity of the
Boolean function $HAM_n^{n/2}$, where
\[HAM^t_n(x,y)=1 \iff dist(x,y)\neq t\iff \sum_i(x_i\oplus y_i)\neq t,\]
for binary strings $x,y$ of length $n$ and the Hamming distance $dist$.
We then show, using methods of de Wolf \cite{W00}, that the nondeterministic (i.e.,
one-sided unbounded error) quantum
communication complexity of $HAM_n^{n/2}$ is $O(\log n)$. So we get an
exponential gap between the nondeterministic quantum and bounded error
quantum complexities.
Since it is also known that the equality function $EQ_n$ has (classical)
bounded error protocols with $O(\log n)$ communication \cite{KN97}, while its
nondeterministic quantum communication complexity is $\Theta(n)$
\cite{W00}, we get the following separation.\footnote{Let $BQP$ denote
  the bounded error quantum communication complexity, $NQP$ the
  nondeterministic quantum
communication complexity, $QC$ the weakly unbounded error quantum
communication complexity (see section 2.2 for definitions).}

\begin{corollary}
There are total Boolean functions $HAM_n^{n/2},EQ_n$ on $2n$
 inputs each, such that
\begin{remunerate}
\item $NQC(HAM_n^{n/2})=O(\log n)$ and $BQC(HAM_n^{n/2})=\Omega(n/\log n)$,
\item $BQC(EQ_n)=O(\log n)$ and $NQC(EQ_n)=\Omega(n)$.
\end{remunerate}
\end{corollary}

Furthermore we give quite tight lower and upper bounds for $HAM_n^t$ for general values of $t$, establishing that bounded error quantum communication does not give a significant speedup compared to classical bounded error communication
for these problems.

We then turn to several other techniques for proving lower bounds,
which are also based on the Fourier transform. We concentrate on
functions $f(x,y)=g(x\diamond y)$, for
$\diamond\in\{\wedge,\oplus\}$, the bitwise conjunction and parity operators.
We prove that for $\diamond=\wedge$, if we choose any Fourier
coefficient $\hat{g}_z$ of $g$, then $|z|/(1-\log |\hat{g}_z|)$
yields a lower bound on the bounded error quantum
communication complexity of $f$. Averaging over all coefficients
leads to a bound given by the average sensitivity of
$g$ divided by the entropy of the squared Fourier coefficients. We then show
another bound for $\diamond=\oplus$ in terms of the entropy of the
Fourier coefficients and
obtain a result solely in terms of the average sensitivity by combining
both results.

\begin{corollary}
For all functions $f$, so that both
$g(x\wedge y)$ and $g(x\oplus y)$ with $g:\{0,1\}^n\to\{0,1\}$
reduce to $f$:
\[BQC(f)=\Omega\left( \sqrt{\frac{\bar{s}(g)}{\log n} }\right).\]\end{corollary}

If e.g.~$f(x,y,z)=g(x\wedge y\oplus z)$, with $x$ held by Alice and
$y,z$ held by Bob, the required reductions are trivial. For many
functions, e.g.~$g=MAJ_n$, it is easy to reduce
$g(x\oplus y)$ on $2\cdot n$ inputs directly to
$g(x\wedge y)$ on more inputs using $x_i\oplus y_i=\neg
x_i\wedge y_i+x_i\wedge \neg y_i$ (plus the addition of a few dummy variables),
and so the lower bound of
corollary 1.2 can sometimes be used for $g(x\wedge y)$.

We then modify the lower bound methods, and show how
we may replace the Fourier
coefficients by the singular values of the communication matrix
(divided by $2^n$). This means that we may replace the Fourier
transform by other unitary transforms and sometimes get much stronger
lower bounds.

Application of the new methods to the Boolean function
\[MAJ_n(x,y)=1\iff \sum_i(x_i\wedge y_i)\ge n/2\]
yields a lower bound
of $\Omega(n/\log n)$ for its bounded error quantum communication
complexity.
$MAJ_n$ is a function, for which neither bounded error quantum nor
nondeterministic quantum protocols are efficient, while the
discrepancy bound is still only $O(\log n)$.

We then apply the same approach to\[COUNT^t_n(x,y)=1\iff
\sum_i(x_i\wedge y_i) = t.\]
These functions have a classical complexity of $\Theta(n)$ for all
$t\le n/2$, since one can easily reduce the disjointness problem to
these functions ($DISJ_n$ is the complement of $COUNT^0_n$).
We show the following:

\begin{corollary}
\[\Omega(n^{1-\epsilon}/\log n)\le
BQC(COUNT^{n^{1-\epsilon}}_n)
\le O(n^{1-\epsilon/2}\log n).\]

\[BPC(COUNT^t_n)=\Theta(n)\mbox{ for all }t\le n/2.\]
\end{corollary}

These are the first lower bounds for functions which allow a
polynomial quantum speedup.

Prior to this paper the only known general method for proving lower bounds for
the bounded error quantum communication complexity has been the
discrepancy method. We show that for any application of the discrepancy bound
to $HAM_n^{t},MAJ_n$, and $COUNT^t_n$, the result is only $O(\log n)$. To
do so we characterize the discrepancy bound within a constant
multiplicative factor and an
additive log-factor as the classical weakly unbounded error
communication complexity $PC$ (see sections 2.2/2.4 for definitions).

\begin{corollary} For all $f:\{0,1\}^n\times\{0,1\}^n\to\{0,1\}:$
\[\max_\mu\log(1/disc_\mu(f))
\le O(PC(f))\le O(\max_\mu\log(1/disc_\mu(f))+\log n).\]
\end{corollary}

This explains why the discrepancy bound is usually in applications
not a good lower bound for bounded error communication complexity,
since the weakly unbounded error complexity is always asymptotically
at most as large
as e.g.~the classical nondeterministic complexity.
For our examples the new lower bound methods are
exponentially better than the discrepancy bound. In the light of
corollary 1.4 it becomes clear that actually lower bounds using
discrepancy follow the approach of simulating quantum bounded error
protocols by classical unbounded error protocols and subsequent
application of a classical lower bound.

We conclude also
that the discrepancy bound subsumes other
methods for proving lower bounds on the weakly unbounded error
communication complexity \cite{DKMW92}.
Furthermore we investigate quantum protocols with weakly unbounded
error and show that quantum and classical weakly
unbounded error communication complexity are asymptotically
equivalent.

The organization of the paper is as follows. In section 2 we describe the
necessary technical background. Section 3 shows how we can decompose
quantum protocols into weighted rectangle covers of the communication
matrix. Sections 4 and 6 then describe our main lower bounds
techniques, while sections 5 and 7 show how to apply these to specific
functions and derive corollaries 1.1 and 1.3.
Section 8 is concerned with the power of classical and
quantum weakly unbounded error protocols. Section 9 discusses recent
developments and open problems.

\section{Preliminaries}

Note that we consider functions with range $\{0,1\}$ as well as
with range $\{-1,1\}$. If a result is stated for functions with range
$\{0,1\}$ then it also holds for $\{-1,1\}$. Some results are stated only for
functions with range $\{-1,1\}$. The communication complexity
does not depend on that choice, so this means that certain parameters in the
lower bounds are dependent on the range.

\subsection{Quantum States and Transformations}

Quantum mechanics is usually formulated in terms of states and
transformations of states. See \cite{NC00} for general information
on this topic with an orientation on quantum computing.

In quantum mechanics pure states are unit norm vectors in a Hilbert
space, usually $\Com^k$. We use the Dirac notation for pure states. So
a pure state is denoted $|\phi\rangle$ or
$\sum_{x\in\{0,\ldots,k-1\}}\alpha_x|x\rangle$ with
$\sum_{x\in\{0,\ldots,k-1\}}|\alpha_x|^2=1$ and with
$\{\,|x\rangle\,|x\in\{0,\ldots,k-1\}\}$ being an orthonormal basis of
$\Com^k$.

Inner products in the Hilbert space are denoted
$\langle\phi|\psi\rangle$.

If $k=2^l$ then the basis is also denoted
$\{\,|x\rangle\,|x\in\{0,1\}^l\}$. In this case the space $\Com^{2^l}$
is the $l$-wise tensor product of the space $\Com^2$. The latter space
is called a qubit, the former space consists of $l$ qubits.

As usual measurements of observables and unitary
transformations are considered as basic operations on states, see
\cite{NC00} for definitions.

\subsection{The Communication Model}

Now we provide definitions of the computational models
considered in the paper. We begin with the model of classical
communication complexity.

\begin{definition}
 Let $f:X\times Y\to\{0,1\}$ be a function. In a
 communication protocol player Alice and Bob receive $x\in X$ and
 $y\in Y$ and compute $f(x,y)$. The players exchange binary encoded messages. The
  communication complexity of a protocol is the worst case number of
  bits exchanged for any input. The deterministic communication
  complexity $DC(f)$ of $f$ is the complexity of an optimal protocol
  for $f$.
  
  In a randomized protocol both players have access to private random
  bits. In the bounded error model the output is required to be
  correct with probability
  $1-\epsilon$ for some constant $1/2>
\epsilon\ge 0$. The bounded error randomized
  communication complexity of a function $BPC_\epsilon(f)$ is then
  defined analogously to the deterministic communication complexity.
  We set $BPC(f)=BPC_{1/3}(f)$.

  In a weakly unbounded error protocol the output has to be correct
  with probability exceeding $1/2$. If the worst case error of the
  protocol (over all inputs) is $1/2-\delta$ and the worst case
  communication is $c$, then the cost of the protocol is defined as
  $c-\lfloor\log\delta\rfloor$.
  The cost of an optimal weakly unbounded error protocol for a
  function is called $PC(f)$.
\end{definition}
  
\begin{definition} Let us note that the {\it communication matrix} of a function
$f:X\times Y\to Z$ is the matrix with rows labeled by $x\in X$,
columns labeled by $y\in Y$, and the entry in row $x$ and column $y$
equal to $f(x,y)\in Z$. A {\it rectangle} in the communication matrix
is a product set of inputs labeled by $A\times B$ with $A\subseteq X$ and $B\subseteq
Y$. Such a rectangle is {\it monochromatic}, iff all its entries are
equal.
\end{definition}

It is easy to see that a deterministic protocol partitions the
communication matrix into a set of monochromatic rectangles, each
corresponding to the set of inputs sharing the same communication
string produced in the run of the protocol.

The above notion of weakly unbounded error protocols coincides
with another type of protocols, namely majority nondeterministic protocols, which accept an input,
whenever there are more nondeterministic computations leading to
acceptance than to rejection. For a proof see theorem 10 in \cite{HR90}. So weakly
unbounded error protocols correspond to certain majority covers for
the communication matrix as follows:

\begin{fact}
There is a weakly unbounded error protocol with cost $O(c)$, iff there is
a set $2^{O(c)}$ rectangles each labeled either 1 or 0, such that for
every input at least
one half of the adjacent rectangles have the label $f(x,y)$.
\end{fact}

Note that there is another type of protocols, truly unbounded error
protocols, in which the cost is not dependent on the error, defined by
Paturi and Simon \cite{PS84}. Recently a linear lower bound for the
unbounded error communication complexity of $IP_n$ has been
obtained in \cite{F01}. It is not hard to see that the same bound
holds for quantum communication as well. An interesting observation is
that
that the lower bound method of \cite{F01} is actually equivalent to
the discrepancy lower bound restricted to the uniform distribution.

Now we turn to quantum communication protocols. For a
more formal definition of quantum protocols see \cite{Y93}.

\begin{definition}
  In a quantum protocol both players have a private set of qubits.
  Some of the qubits are initialized to the input before the start of
  the protocol, the other qubits are in state $|0\rangle$. In a
  communication round one of the players performs some unitary
  transformation on the qubits in his possession and then sends one
  of his qubits to the other player (the latter step does not change
  the global state but rather the possession of individual qubits).
  The choices of the unitary operations and of the qubit to be sent
  are fixed in advance by the protocol.
  
  At the end of the protocol the state of some qubit belonging to one
  player is measured and the result is taken as the output and
  communicated to the other player. The
  communication complexity of the protocol is the number of qubits
  exchanged.
  
  In a (bounded error) quantum protocol the correct answer must be
  given with probability $1-\epsilon$ for some $1/2>\epsilon\ge0$. The
  (bounded error) quantum complexity of a function, called
  $BQC_\epsilon(f)$, is the complexity of an optimal protocol for $f$.
  $BQC(f)=BQC_{1/3}(f)$.

  In a weakly unbounded error quantum protocol the output has to be correct
  with probability exceeding $1/2$. If the worst case error of the
  protocol (over all inputs) is $1/2-\delta$ and the worst case
  communication is $c$, then the cost of the protocol is defined
  as $c-\lfloor\log\delta\rfloor$.
  The cost of an optimal weakly unbounded error protocol for a
  function is called $QC(f)$.
 
  In a nondeterministic quantum protocol for a Boolean function $f$
  all inputs in $f^{-1}(0)$ have to be rejected with certainty, while
  all other inputs have to be accepted with positive probability. The
  corresponding complexity is denoted $NQC(f)$.
\end{definition}

We have to note that in the defined model no intermediate measurements
are allowed to control the choice of qubits to be sent or the time of
the final measurement. Thus for all inputs the same amount of
communication and the same number of message exchanges is used. As a
generalization one could allow
intermediate measurements, whose results could be used to choose 
(several) qubits to be sent and possibly when to stop the communication
protocol. One would have to make sure that the receiving player knows
when a message ends. While the model in our definition is in the
spirit of the ``interacting quantum circuits'' definition given by Yao \cite{Y93},
the latter definition would more resemble ``interacting quantum
Turingmachines''.
Obviously the latter model can be simulated by the former such that
in each communication round exactly one qubit is communicated. All
measurements can then be deferred to the end by standard techniques. This
increases the overall communication by a factor of 2 (and the number
of message exchanges by a lot).

\subsection{Fourier Analysis}

We consider functions $f:\{0,1\}^n\to\R$.
Define \[\langle
f,g\rangle=\frac{1}{2^n}\sum_{x\in\{0,1\}^n} f(x)\cdot g(x) \]
as inner product and use the norm $||f||_2=\sqrt{\langle f,f\rangle}.$
We identify $\{0,1\}^n$ with $\Z_2^n$ and describe the
Fourier transform. A basis for the space of functions
from $\Z_2^n\to\R$ is given by \[\chi_z(x)=(-1)^{IP_n(x,z)}\] for all $z\in\Z_2^n$.
Then the Fourier transform of $f$ with respect to that basis is
\[\sum_z \hat{f}_z \chi_z,\] where the $\hat{f}_z=\langle f,
\chi_z\rangle$ are called the
Fourier coefficients of $f$. If the functions are viewed as vectors,
this is closely related to the Hadamard transform used in quantum
computing.

The following facts are well-known.

\begin{fact}[Parseval]
For all $f$: $||f||_2^2=\sum_z \hat{f}_z^2$.
\end{fact}

\begin{fact}[Cauchy-Schwartz]
\[ \sum_z\hat{f}_z^2\cdot\sum_z\hat{g}_z^2\ge\left(\sum_z
  |\hat{f}_z\cdot \hat{g}_z|\right)^2.\]
\end{fact}

When we consider (communication) functions $f:\Z_2^n\times
\Z_2^n\to\R$, we use the basis functions

\[\chi_{z,z'}(x,x')=(-1)^{IP_n(x,z)+IP_n(x',z')}\] for all
$z,z'\in\Z_2^n\times \Z_2^n$ in Fourier transforms.
The Fourier transform of $f$ with respect to that basis is
\[\sum_{z,z'} \hat{f}_{z,z'} \chi_{z,z'},\] where the
$\hat{f}_{z,z'}=\langle f, \chi_{z,z'}\rangle$ are the Fourier coefficients of $f$.

We will decompose communication protocols into sets of weighted
rectangles. For each rectangle $R_i=A_i\times B_i\subseteq \{0,1\}^n\times
\{0,1\}^n$ let $R_i,A_i,B_i$ also denote the characteristic functions associated
to the rectangle. Then let $\alpha_i=|A_i|/2^n$ be the uniform
probability of $x$ being in the rectangle, and $\beta_i=|B_i|/2^n$ be the uniform
probability of $y$ being in the rectangle. Let $\hat{\alpha}_{z,i}$
denote the Fourier coefficients of $A_i$ and $\hat{\beta}_{z,i}$ the Fourier
coefficients of $B_i$. It is easy to see that $\hat{\alpha}_{z,i}\cdot
\hat{\beta}_{z',i}$ is the $z,z'$-Fourier coefficient of the rectangle
function $R_i$.

For technical reasons we will sometimes work with functions $f$, whose range
is $\{-1,1\}$. Note that we can set $f=2g-1$ for a function $g$ with
range $\{0,1\}$. Since the Fourier transform is linear, the effect on
the Fourier coefficients is that they get multiplied by 2 except for the
coefficient of the constant basis function, which is also decreased by 1.

\subsection{Discrepancy, Sensitivity, and Entropy}

We now define the discrepancy bound.
\begin{definition} Let $\mu$ be any distribution on
 $\{0,1\}^n\times\{0,1\}^n$ and $f$ be any function
 $f:\{0,1\}^n\times\{0,1\}^n\to\{0,1\}$. 
Then let \[disc_\mu(f)=\max_R|\mu(R\cap f^{-1}(0))-\mu(R\cap f^{-1}(1))|,\]
where $R$ runs over all rectangles in the communication matrix of
$f$.

Then denote $disc(f)=\min_\mu disc_\mu (f).$
\end{definition}

The application to communication complexity is as follows
(see \cite{K95} for a less general statement, we also provide a proof
for completeness at the end of section 3):

\begin{fact} For all $f$:
\[BQC_{1/2-\epsilon}(f)=\Omega(\log(\epsilon/disc(f))).\]

A quantum protocol which computes a function $f$ correctly with probability
$1/2+\epsilon$ over a distribution $\mu$ on the inputs (and over its
measurements) needs at least
$\Omega(\log(\epsilon/disc_\mu(f)))$ communication.
\end{fact}

We will prove a lower bound on quantum communication complexity in
terms of average sensitivity.
The average sensitivity of a function measures how many of the $n$
possible bit flips in a random input change the function value. We
define this formally for functions with range $\{-1,1\}$.

\begin{definition} Let $f:\{0,1\}^n \to\{-1,1\}$ be a function.
For $a\in\{0,1\}^n$ let
$s_a(f)=\sum_{i=1}^n\frac{1}{2}|f (a)-f(a\oplus e_i )|$ for the vector
$e_i$ containing a one at position $i$ and zeroes elsewhere. $s_a(f)$
is the sensitivity of $f$ at $a$. Then the average sensitivity of $f$
is defined $\bar{s}(f)=\sum_{a\in\{0,1\}^n}\frac{1}{2^n} s_a(f)$.
\end{definition}

The connection to Fourier analysis is made by the following fact first
observed in \cite{KKL88}.

\begin{fact} For all $f:\{0,1\}^n \to\{-1,1\}:$
\[\bar{s}(f)=\sum_{z\in\{0,1\}^n}|z|\cdot\hat{f}_z^2.\]
\end{fact}

So the average sensitivity can be expressed in terms of the expected
``height'' of Fourier coefficients under the distribution induced by
the squared coefficients.

One more notion we will use in lower bounds is entropy.

\begin{definition}
The entropy of a vector $(f_1,\ldots,f_m)$ with $f_i\ge 0$ for all $i$
and $\sum f_i\le 1$ is $H(f)=-\sum_{i=1}^m
f_i\log f_i.$\end{definition}

We follow the convention $0\log 0=0$.
We will consider the entropy of the vector of squared
Fourier coefficients $H(\hat{f}^2)=-\sum_z \hat{f}_z^2\log(\hat{f}_z^2).$
This quantity has the following useful property.

\begin{lemma}
For any $f:\{0,1\}^n\to\R$ with $||f||_2\le 1:$
\[H(\hat{f}^2)\le 2\log\left( 1+\sum_{z\in\{0,1\}^n} |\hat{f}_z|\right).\]
\end{lemma}

\begin{proof}
\begin{eqnarray*}
H(\hat{f}^2)&=&\sum_z \hat{f}^2_z\log\frac{1}{|\hat{f}_z|^2}\\
&=&2\left(\sum_z \hat{f}^2_z\log\frac{1}{|\hat{f}_z|}+(1-\sum_z
  \hat{f}^2_z)\cdot \log 1\right)\\
&\le& 2\log\left(\sum_z
\hat{f}^2_z\frac{1}{|\hat{f}_z|}+(1-\sum_z
\hat{f}^2_z)\cdot 1\right) \hspace{.5cm}\mbox{ by Jensen's inequality}\\
&\le&2\log\left(1+\sum_z|\hat{f}_z|\right).\end{eqnarray*}
\qquad\end{proof}

\section{Decomposing Quantum Protocols}

In this section we show how to decompose a quantum protocol into a
set of weighted rectangles, whose sum approximates the communication
matrix.

\begin{lemma}
For all Boolean functions
$f:\{0,1\}^n\times\{0,1\}^n\to\{0,1\}$, and for all constants
$1/2>\epsilon>0$:

If there is a quantum protocol for $f$ with communication $c$
and error 1/3,

then there is a real $\alpha\in[0,1]$, and a set of $2^{O(c)}$
rectangles $R_i$ with weights
$w_i\in\{-\alpha,\alpha\}$, so that
\[\sum_i w_i R_i[x,y] \in
\left\{
\begin{array}{ll}
[1-\epsilon,1]& \mbox{ for } f(x,y)=1\\

[0,\epsilon]&  \mbox{ for } f(x,y)=0.\\
\end{array}
\right.
\]
\end{lemma}

\begin{proof}
First we perform the usual success amplification to boost the success
probability of the quantum protocol to $1-\epsilon/4$, increasing the
communication to $O(c)$ at most, since $\epsilon$ is assumed to be a
constant.
Using standard techniques \cite{BV97} we can assume that all amplitudes used in
the protocol are real. Now we employ the following fact proved in
\cite{K95} and \cite{Y93}.

\begin{fact}
The final state of a quantum protocol exchanging $c$ qubits on an
input $(x,y)$ can be written
\[\sum_{m\in\{0,1\}^c} \alpha_m(x)\beta_m(y)
|A_m(x)\rangle|m_c\rangle|B_m(y)\rangle,\]
where $|A_m(x)\rangle,|B_m(y)\rangle$ are pure states and
$\alpha_m(x),\beta_m(y)$ are real numbers from the interval $[-1,1]$.
\end{fact}

Now let the final state of the protocol on $(x,y)$ be
\[\sum_{m\in\{0,1\}^c} \alpha_m(x)\beta_m(y)
|A_m(x)|m_c\rangle|B_m(y)\rangle,\] and let
$\phi(x,y)=$\[\sum_{m\in\{0,1\}^{c-1}} \alpha_{m1}(x)\beta_{m1}(y)
|A_{m1}(x)\rangle |1\rangle|B_{m1}(y)\rangle\] be the part of the state which yields
output 1. The acceptance probability of the protocol on $(x,y)$ is now
the inner product $\langle\phi(x,y)|\phi(x,y)\rangle$.
Using the convention
\[a_{mp}(x)=\alpha_{m1}(x)\alpha_{p1}(x)\langle A_{m1}(x)|A_{p1}(x) \rangle,\]
\[b_{mp}(y)=\beta_{m1}(y)\beta_{p1}(y)\langle B_{m1}(y)|B_{p1}(y) \rangle,\]
this can be written as $\sum_{m,p}a_{mp}(x)b_{mp}(y)$. Viewing
$a_{mp}$ and $b_{mp}$ as $2^n$-dimensional vectors, and summing their
outer products over all $m,p$ yields a sum
of $2^{O(c)}$ rank 1 matrices containing reals between -1 and
1. Rewrite this sum as $\sum_i \alpha_i\beta_i^T$ with $1\le i\le
2^{O(c)}$ to save notation.
The resulting matrix is an approximation of the communication matrix
within component\-wise error $\epsilon/4$.

In the next step define for all $i$ a set $P_{\alpha,i}$ of the indices of positive
entries in
$\alpha_i$, and the set $N_{\alpha,i}$ of the indices of negative entries of
$\alpha_i$. Define $P_{\beta,i}$ and $N_{\beta,i}$ analogously. We
want to have that all rank 1 matrices either have only positive or
only negative entries. For this we split the matrices into 4 matrices
each, depending on the positivity/negativity of $\alpha_i$ and
$\beta_i$. Let \[\alpha'_i(x)=\left\{\begin{array}{ll}
0& \mbox{ if } x\in N_{\alpha,i}\\
\alpha_i(x)& \mbox{ if } x\in P_{\alpha,i}
\end{array}\right.,\]
and analogously for $\beta'_i$, then set the positive entries in
$\alpha_i$ and $\beta_i$ to 0. Consider the sum
$\sum_i (\alpha_i\beta_i^T) + \sum_i (\alpha'_i\beta_i^T) + \sum_i
(\alpha_i\beta_i'^T) + \sum_i (\alpha_i'\beta_i'^T).$ This sum 
equals the previous sum, but here all matrices are either nonnegative
or nonpositive.
Again rename the indices so that the sum is written
$\sum_i \alpha_i\beta_i^T$ (to save notation).

At this point we have a set of rank one matrices which are either
nonnegative or nonpositive with the above properties.
We want to round entries and split matrices into uniformly weighted
matrices. Let $C$ denote the number of matrices used until now.

Consider the intervals $[0,\epsilon/(16C)\,\,]$, and
$[\epsilon/(16C)\cdot k, \epsilon/(16C)\cdot (k+1)\,\,]$,
for all $k$ up to the least $k$, so that the last
interval includes 1. Obviously there are $O(C)$ such intervals.
Round every positive $\alpha_i(x)$ and $\beta_i(x)$ to the upper bound
of the first interval it is included in, and change the negative
entries analogously by rounding to the
upper bounds of the corresponding negative intervals.
The overall error introduced on an input $(x,y)$ in the
approximating sum $\sum_i \alpha_i(x)\beta_i(y)$ is at most
\begin{eqnarray*}
&&\sum_i \alpha_i(x)\cdot\epsilon/(16C)\\&&+\sum_i\beta_i(y)\cdot\epsilon/(16C)+C\cdot
\epsilon^2/(16C)^2\\
&\le&\epsilon/4.\end{eqnarray*}
The sum of the matrices is now between
$1-\epsilon/2$ and $1+\epsilon/4$ for inputs in $f^{-1}(1)$ and
between $-\epsilon/4$ and $\epsilon/2$ for inputs in $f^{-1}(0)$. Add
a rectangle with weight $\epsilon/4$ covering all inputs.
Dividing all weights by $1+\epsilon/2$ renormalizes again without increasing the
error beyond $\epsilon$.

Now we are left with $C$ rank 1 matrices $\alpha_i\beta_i^T$ containing
entries from a $O(C)$ size set only. Splitting the rank 1
matrices into rectangles containing only the entries with one of the
values yields $O(C^2)$ weighted rectangles, whose (weighted) sum
approximates the communication matrix within error $\epsilon$.

In a last step we replace any rectangle with an absolute weight value of
$\epsilon/(16C(1+\epsilon/2))\cdot k$ by $k$ rectangles with weights $\pm \alpha$ for
$\alpha=\epsilon/(16C(1+\epsilon/2 ))$. The rectangle weighted
$\epsilon/4$ can be replaced by a set of rectangles with weight
$\alpha$ each, introducing negligible error.
\qquad\end{proof}

Now we show the analogous lemma, when we also want to improve the error
probability beyond a constant. If we would simply decrease the error to $1/2^d$
by repeating the protocol before constructing the cover, then we would
be forced to work with high precision in all steps, increasing the
size of the cover to $2^{O((c+d)d)}$, which is undesirable for large $d$.
Instead we first construct a cover with constant error as before and
then improve the quality of the cover directly.

\begin{lemma}
For all Boolean functions
$f:\{0,1\}^n\times\{0,1\}^n\to\{0,1\}$, and for all $d\ge 1$:

If there is a quantum protocol for $f$ with communication $c$
and error 1/3,

then there is a real $\alpha\in[0,1]$, and a set of $2^{O(dc)}$
rectangles $R_i$ with weights
$w_i\in\{-\alpha,\alpha\}$, so that
\[\sum_i w_i R_i[x,y] \in
\left\{
\begin{array}{ll}
[1-1/2^d,1]& \mbox{ for } f(x,y)=1\\

[0,1/2^d]&  \mbox{ for } f(x,y)=0.\\
\end{array}
\right.
\]
\end{lemma}

\begin{proof}
We start with the result of the previous lemma. The obtained set of
rectangles approximates the communication matrix within error
$\epsilon$ for some small constant $\epsilon.$ Call these
rectangles $R_i$ and their weights $\alpha_i=\pm \alpha$.

Doing the same construction for the rejecting part of the final state
of the original protocol we get
a set of $2^{O(c)}$ weighted rectangles, such that the sum of these is
between 0 and $\epsilon$ on every $x,y\in f^{-1}(1)$ and between
$1-\epsilon$ and 1 for every $x,y\in f^{-1}(0)$. Call these
rectangles $R'_i$. Due to the previous construction their weights can
be assumed to be also
$\alpha_i'=\pm \alpha$. Note that for all $x,y:$
\[\sum_i (\alpha_i R_i(x,y))+\sum_i (\alpha'_i R'_i(x,y))\le1.\]

We construct our new set of rectangles as follows.
For every ordered $k$ tuple of rectangles containing at least $k/2$
rectangles $R_i$ and at most $k/2$ rectangles $R_i'$ we form a new
rectangle by intersecting all of the rectangles in the tuple. The
weight of the new
rectangle is the product of the weights of its constituting
rectangles. Now we consider the sum of all rectangles obtained this
way.

The number of new rectangles is at most $2^{O(ck)}$.
The sum of the weights of rectangles adjacent to a
zero input $x,y$ of the function is
\begin{eqnarray*}
&\sum_{j\le k/2} {k\choose j} \cdot (\sum_i \alpha_i R_i(x,y))^{k-j}\cdot
(\sum_i \alpha'_i R'_i(x,y))^{j}\\
\le & \sum_{j\le k/2} {k\choose j} \cdot \epsilon_{x,y}^{k-j} \cdot
(1-\epsilon_{x,y})^{j}\\
\le & 2^{-\Omega(k)}\\
\end{eqnarray*}
for some $\epsilon>\epsilon_{x,y}>0$ (see e.g.~lemma 2.3.5 in
\cite{G90} for the last inequality).
The same sum of weights is also clearly at least 0.
The sum of the weights of rectangles adjacent to a
one input $x,y$ of the function is
\begin{eqnarray*}
&\sum_{j\le k/2} {k\choose j} \cdot (\sum_i \alpha_i R_i(x,y))^{k-j}\cdot
(\sum_i \alpha'_i R'_i(x,y))^{j}\\
\ge & \sum_{j\le k} {k\choose j} \cdot (1-\epsilon_{x,y})^{k-j} \cdot
\epsilon_{x,y}^{j}-
\sum_{j\le k/2} {k\choose j} \cdot \epsilon_{x,y}^{k-j} \cdot
(1-\epsilon_{x,y})^{j}\\
\ge & 1-2^{-\Omega(k)}\\
\end{eqnarray*}
for some $\epsilon>\epsilon_{x,y}>0$.
The same sum of weights is also clearly at most 1.

So choosing $k=\Theta(d)$ large enough yields the desired set of
rectangles.
\qquad\end{proof}

At first glance the covers obtained in this section seem to
be very similar to majority covers: we have a set of rectangles with
either negative or positive weights of absolute value $\alpha$,
and if the weighted sum of rectangles adjacent to some input exceeds a threshold,
then it is a 1-input. But we
have one more property, namely that summing the weights of the
adjacent rectangles approximates the function value. Actually the lower
bounds in the next sections and the characterization of majority covers
(and weakly unbounded error protocols and the discrepancy bound) in
section 8 show that there is an exponential difference between the
sizes of the two types of covers.

Now we state another form of the lemma, this time if the error is
close to 1/2, the proof is essentially the same as for lemma 3.1.

\begin{lemma}
For all Boolean functions
$f:\{0,1\}^n\times\{0,1\}^n\to\{0,1\}$, and for all $1/2>\epsilon>0$:

If there is a quantum protocol for $f$ with communication $c$
and error $1/2-\epsilon$,

then there is a real $\alpha\in[0,1]$, and a set of
$2^{O(c)}/\epsilon$ rectangles $R_i$ with weights
$w_i\in\{-\alpha,\alpha\}$, so that
\[\sum_i w_i R_i[x,y] \in
\left\{
\begin{array}{ll}
[1/2+\epsilon/2,1]& \mbox{ for } f(x,y)=1\\

[0,1/2-\epsilon/2]& \mbox{ for } f(x,y)=0.
\end{array}
\right.
\]
\end{lemma}

Note that all results of this section easily generalize to functions with
range $\{-1,+1\}$. Furthermore all the results generalize to partial
functions, i.e., the functions may be undefined on some inputs. For
those inputs the weighted covers produce a arbitrary weight between 0
and 1.

As an application of the decomposition results we now prove fact 2.8.
A proof of this result seems to be available only in the thesis of Kremer
\cite{K95} and is stated in less generality there, so we include a proof here.

{\em Proof of fact 2.8.} Obviously it suffices to prove the second
statement. Let $\mu$ be any distribution on the inputs. 
Assume there is a protocol with communication $c$ so that the average
correctness probability over $\mu$ and the measurements of the
protocol is at least $1/2+\epsilon$.

Let $P(x,y)$ denote the probability that the protocol accepts $x,y$
and $K(x,y)$ denote the probability that the protocol is correct on $x,y$.
W.l.o.g.~we assume that $\mu(f^{-1}(1))\ge\mu(f^{-1}(0))$.
Then we have
\begin{eqnarray*}
&&\sum_{x,y\in f^{-1}(1)} \mu(x,y)P(x,y)\\&&-\sum_{x,y\in f^{-1}(0)}
\mu(x,y) P(x,y)\\
&=&\sum_{x,y\in f^{-1}(1)} \mu(x,y)K(x,y)\\&&+\sum_{x,y\in f^{-1}(0)}
\mu(x,y) K(x,y) -\mu(f^{-1}(0))\\
&\ge& 1/2+\epsilon-1/2=\epsilon.\end{eqnarray*}

Following the construction of lemma 3.4 we get a set of
$C=2^{O(c)}/\epsilon$ rectangles $R_i$ with weights $w_i$ so that the sum of these
approximates the acceptance probability of the protocol with
componentwise additive error $\epsilon/2$.
Then
\begin{eqnarray*} &&\sum_{x,y\in f^{-1}(1)}
\mu(x,y) \sum_{1\le i\le C} w_i R_i(x,y)\\
&-&\sum_{x,y\in f^{-1}(0)}
\mu(x,y) \sum_{1\le i\le C} w_i R_i(x,y) \ge
\epsilon-\epsilon/2.\end{eqnarray*}

Exchanging sums gives us
\begin{eqnarray*}
\sum_{1\le i\le C} w_i \left( \sum_{x,y\in f^{-1}(1)}
\mu(x,y) R_i(x,y)-\right.\\
\left. \sum_{x,y\in f^{-1}(0)}
\mu(x,y) R_i(x,y) \right)\ge \epsilon/2\end{eqnarray*}
and
\[\sum_{1\le i\le C} w_i (\mu( f^{-1}(1)\cap R_i )-\mu( f^{-1}(0)\cap
R_i))\ge \epsilon/2.\]

Thus there is a rectangle $R_i$ with $\mu( f^{-1}(1)\cap R_i )-\mu( f^{-1}(0)\cap
R_i)\ge (\epsilon/2)/C,$ but for all rectangles we have $\mu(
f^{-1}(1)\cap R_i )-\mu( f^{-1}(0)\cap
R_i)\le disc_\mu(f)$, hence $disc_\mu(f)\ge (\epsilon/2)/C$ and finally
\[\frac{2^{O(c)}}{\epsilon}=C\ge (\epsilon/2 )/disc_\mu(f)\Rightarrow
c\ge\Omega\left(\log\frac{\epsilon}{disc_\mu(f)} \right).\] \qed

\section{A Fourier Bound}

In this section we describe a lower bound method first developed by
Raz \cite{R95} for classical bounded error communication complexity.
We prove that the same method is applicable in the
quantum case, using the decomposition results from the previous
section. The lower bound method is based on the Fourier transform of
the function.

As in section 2.3 we consider the Fourier transform of a
communication function. The basis functions are labeled by pairs of
strings $(z,z')$. Denote by $V$ the set of all pairs $(z,z)$.
Let $E\subseteq V$ denote some subset of indices of Fourier
coefficients.

The basic idea of the lower bound is that the communication must be
large, when the sum of the absolute values of a small set of Fourier
coefficients is large.

\begin{theorem}
Let $f$ be a total Boolean function
$f:\{0,1\}^n\times\{0,1\}^n\to\{0,1\}$.

Let $E\subseteq V$.
Denote $\kappa_0=|E|$ (the number of coefficients considered)
and $\kappa_1=\sum_{(z,z)\in E} |\hat{f}_{z,z}|$ (the absolute value sum
of coefficients considered). Then:

If $\kappa_1\ge\Omega(\sqrt{\kappa_0})$, then $BQC(f)=\Omega(\log (\kappa_1))$.

If $\kappa_1\le O(\sqrt{\kappa_0})$, then
$BQC(f)=\Omega(\log (\kappa_1) / (\log(\sqrt{\kappa_0})-\log(\kappa_1)+1))$.
\end{theorem}

\begin{proof}
We are given any quantum protocol for $f$ with error 1/3 and some
worst case communication $c$. We have to put the stated lower bound
on $c$. Following lemma 3.3 we can find a set of $2^{O(cd)}$ weighted
rectangles, so that the sum of these approximates the communication
matrix up to error $ 1/2^{d}$ for any $d\ge 1$, where the weights are
either $\alpha$, or $-\alpha$ for some real $\alpha$ between 0 and 1.
We will fix $d$ later. Let $\{(R_i,w_i)|1\le i\le
2^{O(cd)}\}$ denote that set.
Furthermore let $g(x,y)$ denote the function that maps $(x,y)$ to
$\sum_i w_i R_i(x,y)$.

First we give a lower bound on the sum of absolute values of the Fourier
coefficients in $E$ for $g$, in terms of the respective sum for $f$,
using the fact that $g$ approximates $f$. Obviously $||f-g||_2\le
1/2^d$. The identity of Parseval then gives us
\[\sum_{(z,z)\in E} (\hat{f}_{z,z}-\hat{g}_{z,z})^2\le ||f-g||_2^2\le 2^{-2d}.\]

We make use of the following simple consequence of fact 2.6.
\begin{fact}
Let $|||v|||_2=\sqrt{\sum_{i=1}^m v_i^2}$, and $|||v|||_1=\sum_{i=1}^m
|v_i|$.
Then $|||v-w|||_2\ge |||v-w|||_1/\sqrt{m}\ge (|||v|||_1-|||w|||_1)/\sqrt{m}.$
\end{fact}

Hence \begin{eqnarray*}
\sum_E |\hat{g}_{z,z}|&\ge& \sum_E
|\hat{f}_{z,z}|-\sqrt{|E|\cdot\sum_E(\hat{f}_{z,z}-\hat{g}_{z,z})^2}\\
&\ge& \kappa_1-\sqrt{\kappa_0}\cdot 2^{-d}.
\end{eqnarray*}

Thus the sum of absolute values of the chosen Fourier coefficients of
$g$ must be large, if there are not too many such coefficients, or if
the error is small enough to suppress their number in the above
expression.
Call $P=(\kappa_1-\sqrt{\kappa_0}\cdot 2^{-d})$, so $\sum_E
|\hat{g}_{z,z}|\ge P$.

Now due to the decomposition of the quantum protocol used to obtain
$g$, the function is the weighted sum of $C=2^{O(cd)}$ rectangles.
Since the Fourier transform is a linear transformation, the Fourier
coefficients of $g$
are weighted sums of the Fourier coefficients of the rectangles.
Furthermore the Fourier coefficients of a rectangle are the products of the
Fourier coefficients of the characteristic functions of the sets constituting
the rectangle, as argued in section 2.3.
So $\hat{g}_{z,z}=\sum_i w_i\cdot \hat{\alpha}_{z,i}\cdot\hat{\beta}_{z,i}$ and
\begin{equation}\sum_E|\hat{g}_{z,z}|\le\sum_E\sum_i |w_i\cdot
 \hat{\alpha}_{z,i}\cdot \hat{\beta}_{z,i}|.\end{equation}

For all rectangles $R_i$ we have $\sum_E\hat{\alpha}_{z,i}^2\le
||A_i||_2^2\le 1$ by the identity of Parseval. Using the
Cauchy-Schwartz inequality (fact 2.6) we get
$\sum_E|\hat{\alpha}_{z,i}\hat{\beta}_{z,i}|\le1.$ But according to
(4.1) the
weighted sum of these values, with weights between -1 and 1, adds up to
at least $ P$, and so at least $C\ge P$ rectangles are there, thus
$cd=\Omega(\log P)$.

If now $\kappa_1\ge\Omega(\sqrt{\kappa_0})$, then let $d=O(1)$, and we get the
lower bound $c=\Omega(\log (\kappa_1))$. Otherwise set $d=O(\log
\sqrt{\kappa_0}-\log \kappa_1+1)$ to get $P=\kappa_1/2$ as well as
$c=\Omega(\log (P)/d)=
\Omega(\log (\kappa_1)/
( \log(\sqrt{\kappa_0}) - \log (\kappa_1) +1) )$.
\qquad\end{proof}

Let us note one lemma that is implicit in the above proof,
and which will be used later.

\begin{lemma}
Let $g:\{0,1\}^n\times\{0,1\}^n\to [-1,1]$ be any function such that
there is a set of $Q$ rectangles $R_i$ with weights $w_i\in[-1,1]$ so that
$g(x,y)=\sum_{i=1}^Q w_i R_i(x,y)$ for all $x,y$. Then
\[\sum_{z\in\{0,1\}^n}|\hat{g}_{z,z}|\le Q.\]\end{lemma}

\section{Applications}

In this section we give applications of the lower bound method.

\subsection{Quantum Nondeterminism versus Bounded Error}

We first use the lower bound method to prove that
nondeterministic quantum protocols may be exponentially more efficient than
bounded error quantum protocols. Raz has shown the following \cite{R95}:

\begin{fact}
For the function $HAM_n^{n/2}$ consider the set of Fourier coefficients with
labels from a set $E$ containing those strings $z,z$ with $z$ having
$n/2$ ones. Then
\[\kappa_0={n\choose n/2}, \kappa_1={n\choose n/2}{n/2\choose
 n/4}\frac{1}{2^n}.\]
\end{fact}

Thus $\log(\sqrt{\kappa_0})-\log(\kappa_1)=O(\log n)$.
Also $\kappa_1=\Theta(2^{n/2} /n)$ and thus $\log \kappa_1=\Theta(n)$.

Applying the lower bound method we get

\begin{theorem}
$BQC(HAM_n^{n/2})=\Omega(n/\log n)$.
\end{theorem}

Now we prove that the nondeterministic quantum complexity of $HAM_n^{n/2}$
is small. We use the following technique by de Wolf \cite{W00,HW02}.

\begin{fact}
Let the nondeterministic rank of a Boolean function $f$ be the minimum
rank of a matrix that contains 0 at positions corresponding to inputs
$(x,y)$ with $f(x,y)=0$ and nonzero reals elsewhere. Then $NQC(f)=\log nrank(f)+1$.
\end{fact}

\begin{theorem}
$NQC(HAM_n^{n/2})=O(\log n)$.
\end{theorem}

\begin{proof}
It suffices to prove that the nondeterministic rank is polynomial. 
Define rectangles $M_i$, which include inputs with $x_i=1$ and $y_i=0$, and $N_i$,
which include inputs with $x_i=0$ and $y_i=1$. Let $E$ denote the all
one matrix. Then let $M=\sum_i
(M_i+N_i) - n/2\cdot E$. This is a matrix which is 0 exactly at those inputs
with $\sum_i (x_i\oplus y_i)=n/2$. Furthermore $M$ is composed of
$2n+1$ weighted rectangles and thus the nondeterministic rank of
$HAM_n^{n/2}$ is $O(n)$.
\qquad\end{proof}

\subsection{The Complexity of the Hamming Distance Problem}

Now we determine the complexity of $HAM_n^t$, and show that quantum bounded
error communication does not allow a significant speedup.

\begin{theorem} Let $t:\N\to\N$ be any monotone increasing function
with $t(n)\le n/2$. Then
\[BQC(HAM^{t(n)}_n)\ge \Omega\left(\frac{t(n)}{\log t(n)}+\log
 n\right ).\]
\end{theorem}

\begin{proof}
We already know that the complexity of $HAM^{n/2}_n$ is
$\Omega(n/\log n)$.
Now consider functions $HAM^t_n$ for smaller $t$. The logarithmic
lower bound is obvious from the at most exponential speedup obtainable
by quantum protocols \cite{K95}.

Fixing $n-2t$ pairs of inputs variables to the same values
leaves us with $2t$ pairs of free variables and
the function accepts if $HAM^t_{2t}$ accepts on these inputs. Thus
the lower bound follows.
\qquad\end{proof}

\begin{theorem}
\[BPC(HAM_n^t)=O(t\log n).\]
\end{theorem}

\begin{proof}
  The protocol determines (and removes) positions in which $x,y$ are
different, until no more such positions are present, or until $t+1$
such positions are found, in both
cases the function value can be decided.

Nisan \cite{N93} has given a protocol in which Alice and Bob, given
$n$-bit strings $x,y$, compute the leftmost bit in which $x,y$ differ. The
protocol needs communication $O(\log n-\log\epsilon)$ to solve this problem
with error $\epsilon$. Hence we can find such a position with error $1/(3t)$
and communication $O(\log n)$, since $t\le n$. So Alice and Bob can
determine with error 1/3, whether there are exactly $t$ differences between $x$ and
$y$, using communication $O(t\log n)$ as claimed.
\qquad\end{proof}

\section{More Fourier Bounds}

In this section we develop more methods for proving lower bounds on quantum
communication complexity in terms of properties of their Fourier
coefficients. Combining them yields a bound in terms of average
sensitivity.

\subsection{A Bound Employing One Fourier Coefficient}

Consider functions of the type $f(x,y)=g(x\wedge y)$. The Fourier
coefficients of
$g$ measure how well the parity function on a certain set of variables
is approximated by $g$.
But if $g$ is correlated with a parity (hopefully on a large set of variables), then $f$
should be correlated with an inner product function.
The hardness result stated in fact 2.8 then gives the intuition for the
first bound of this section.

\begin{theorem}
  For all total functions $f:\{0,1\}^n\times\{0,1\}^n\to\{0,1\}$ with
$f(x,y)=g(x\wedge y)$ and all $z\in\{0,1\}^n:$
\[BQC(f)=\Omega\left(\frac{|z|}{1-\log|\hat{g}_z|}\right).\]\end{theorem}

\begin{proof}
We prove the bound for $g$ with range $\{-1,1\}$. Obviously the bound
itself changes only by a constant factor with this change and the
communication complexity is unchanged.

Let $z$ be the index of any Fourier coefficient of $g$. Let
$|z|=m$. Basically $\hat{g}_z$ measures how well $g$ approximates
$\chi_z$, the parity function on the $m$ variables
which are 1 in $z$. Consider the following distribution $\mu_m$ on
$\{0,1\}^m\times\{0,1\}^m$:
Each variable is set to one with
probability $\sqrt{1/2}$ and to zero with probability
$1-\sqrt{1/2}$. Then every $x_i\wedge y_i$ is one resp.~zero with
probability $1/2$. So under this distribution on the inputs $(x,y)$ to
$f$ we get the uniform distribution on the inputs $z=x\wedge y$ to $g$.

We will get an approximation of $IP_m$
under $\mu_m$ with error $1/2-|\hat{g}_z|/4$ by taking the outputs of a
protocol for $f$ under a suitable distribution.
We then use a hardness result for $IP_m$ given by the following lemma.

\begin{lemma}
Let $\mu_m$ be the distribution on $\{0,1\}^m\times\{0,1\}^m$, that is
the $2m$-wise product of the distribution on $\{0,1\}$, in which 1 is
chosen with probability $\sqrt{1/2}$. Then
\[disc_{\mu_m} (IP_m)\le O(2^{-m/4}).\]
\end{lemma}

Clearly with fact 2.8 we get that computing $IP_m$ with error
$1/2-\epsilon$ under the distribution $\mu_m$ needs quantum communication
$\Omega(m/4+\log\epsilon)$.

Let us prove the lemma. Lindsey's lemma (see e.g.~\cite{BFS86}) states the
following.

\begin{fact}
Let $R$ be any rectangle with $a\times b$ entries in the communication
matrix of $IP_m$. Then let \[\left| |R\cap IP_m^{-1}(1)| - |R\cap
IP_m^{-1}(0)| \right|\le\sqrt{ab2^m}.\]
\end{fact}

The above fact allows
to compute the discrepancy of $IP_n$ under the uniform distribution, and will also be helpful for $\mu_m$.

$\mu_m$ is uniform on the subset of all inputs $x,y$ containing $k$
ones. Consider any rectangle $R$.
There are at most ${2m\choose k}$ inputs with exactly $k$ ones in that
rectangle. Furthermore if we intersect the
rectangle containing all inputs $x,y$ containing $i$ ones in $x$ and
$j$ ones in $y$ with $R$ we get a rectangle containing at most ${m\choose
  i}\cdot{m\choose j}\le{2m\choose i+j}$ inputs. In this way $R$
is partitioned into $m^2$ rectangles, on which $\mu_m$ is uniform and Lindsey's
lemma can be applied. Note that we partition the set of inputs with overall
$k$ ones into up to $m$ rectangles.

Let $\alpha=\sqrt{1/2}$. The probability of any input with $k$ ones is
$(1-\alpha)^{2m-k}\cdot\alpha^k$.
We get the following upper bound on discrepancy under
$\mu_m$:
\begin{eqnarray*}
&& \sum_{i,j=0}^{m}\alpha^{i+j}\cdot(1-\alpha)^{2m-i-j}\cdot
\sqrt{ {m\choose i} {m\choose j} 2^m}\\
&\le&
m2^{m/2}\cdot\sum_{k=0}^{2m}
\alpha^k\cdot(1-\alpha)^{2m-k}\cdot\sqrt{{2m\choose k}}\\
&\le &
m2^{m/2}\cdot\sqrt{2m+1}\cdot\sqrt{\sum_{k=0}^{2m}\alpha^{2k}\cdot(1-\alpha)^{4m-2k}\cdot{2m\choose k}}\\
&\le & m\sqrt{2m+1}2^{m/2} (\alpha^2+(1-\alpha)^2)^m\\
&\le&m\sqrt{2m+1}2^{m/2}(2-\sqrt{2})^m\\
&\le & O(2^{-m/4}).\end{eqnarray*}
This concludes the proof of lemma 6.2.

To describe the way we use this hardness result first assume that
the quantum protocol for $f$ is errorless. The Fourier coefficient for $z$
measures the correlation between $g$ and the parity function $\chi_z$
on the variables that are ones in $z$. We first show that $\chi_{1^m}$
can be computed with error $1/2-|\hat{g}_z|/2$ from $g$ (or its complement). To
see this consider
$\hat{g}_z=\langle g,\chi_z\rangle=\sum_a\frac{1}{2^n} g(a)\cdot
\chi_z(a).$
W.l.o.g.~assume that the first $m$ variables of
$z$ are its ones. So we can rewrite to
\[\hat{g}_z=\sum_{b\in\{0,1\}^{n-m}}
\frac{1}{2^{n-m}}
\sum_{a\in\{0,1\}^{m}}\frac{1}{2^m}
g(ab)\cdot\chi_z(ab).\]
Note that $\chi_z$ depends only on the first $m$ variables.
In other words, if we fix a random $b$, the output of $g$ has an expected advantage
of $|\hat{g}_z| $ over a random choice in computing parity on the cube
spanned by the first
$m$ variables. Consequently there must be some $b$ realizing that
advantage. We fix that $b$, and use $g(ab)$ (or $-g(ab)$) to
approximate $\chi_{1^m}$.
The error of this approximation is $1/2-|\hat{g}_z|/2$.

Next we show that $IP_m$ resp.~$\chi_{1^m}(x\wedge
y)=\chi_z((x\wedge y) \circ b)$ is correlated with $g((x\wedge y)\circ b)$
under some distribution.

Let $\mu'_n$ be a distribution resulting from $\mu_n$, if all $x_i$ and
$y_i$ for $i=m+1,\ldots,n$ are fixed so that $x_i\wedge y_i=b_{i-m}$
and all other variables are chosen as for $\mu_n$.
Then
\begin{eqnarray*}
&& |\sum_{(x,y)\in\{0,1\}^{2\cdot n}}
\mu'_n(x,y)\cdot g(x\wedge y)\cdot\chi_z(x\wedge y) |\\
&=& |\sum_{a\in\{0,1\}^m} g(ab)\cdot
\chi_z(ab)\cdot\sum_{x,y:x\wedge y=ab }\mu'_n(x,y) |\\
&=& |\sum_{a\in\{0,1\}^m} g(ab)\cdot \chi_z(ab)\cdot
\frac{1}{2^m}|\ge|\hat{g}_z| .\end{eqnarray*}

Hence computing $f$ on $\mu'_n$ with no error is at least as
hard as computing $IP_m$ on distribution $\mu_m$ with error
$1/2-|\hat{g}_z|/2$, which needs at
least $\Omega(|z|/4+\log|\hat{g}_z|)$ qubits communication due to the
  discrepancy bound.

We assumed previously that $f$ is computed without error. Now assume
the error of a protocol for $f$ is $1/3$. Then reduce the error probability
to $|\hat{g}_z|/4$ by repeating the protocol $d=O(1-\log|\hat{g}_z|)$
times and taking the majority output. Computing $f$ on $\mu'_n$ with
error $|\hat{g}_z|/4$
is at least as hard as computing $IP_m$ on distribution $\mu_m$ with error
$1/2-|\hat{g}_z|/2+|\hat{g}_z|/4$, which needs at
least $\Omega(|z|/4+\log|\hat{g}_z|)$ qubits communication.
The error introduced by the protocol is smaller than the
advantage of the function $f$ in computing $IP_m$.

So a lower bound of $\Omega(|z|/4+\log|\hat{g}_z|)$ holds for the task
of computing $f$ with error $|\hat{g}_z|/4$. This implies a lower bound
of  \[\frac{\Omega(|z|/4+\log|\hat{g}_z|)}{d}=
\Omega\left(\frac{|z|}{1-\log|\hat{g}_z|}\right).\]
for the task of computing $f$ with error 1/3.
\qquad\end{proof}

 Note that the discrepancy of $f$ in the above theorem  may be
  much higher than the discrepancy of $IP_m$ (leading to weak lower
  bounds for $f$), but that $f$
  approximates $IP_m$ well enough to transfer the lower bound known
  for $IP_m$ (which happens to be provable via low discrepancy).

\subsection{A Sensitivity Bound}

A weaker, averaged form of the bound in the above subsection is the following.
\begin{lemma}
For all functions $f:\{0,1\}^n\times\{0,1\}^n\to\{-1,1\}$ with
$f(x,y)=g(x\wedge y):$
\[BQC(f)=\Omega\left(\frac{\bar{s}(g)}{H(\hat{g}^2)+1}\right).\]\end{lemma}

\begin{proof}
First note that $\bar{s}(g)=\sum_z \hat{g}^2_z |z|$ by fact 2.10. So we
can read the bound
\[BQC(f)=\Omega\left(\frac{\sum_z
 \hat{g}^2_z|z|}{\sum_z \hat{g}_z^2(1-2\log|\hat{g}_z|) }\right).\]

The $\hat{g}^2_z$ define a probability distribution on the
$z\in\{0,1\}^n$. If we choose a $z$ randomly then the expected Hamming
weight of $z$ is $\bar{s}(g)$. Also the expectation of $1-2\log|\hat{g}_z|$ is
$1+H(\hat{g}^2)$.
We use the following lemma.
\begin{lemma}
Let $a_1,\ldots,a_m$ be nonnegative and $b_1,\ldots,b_m$ be positive numbers and
let $p_1,\ldots, p_m$ be a probability distribution.
Then there is an $i$ with:
\[\frac{a_i}{b_i}\ge\frac{\sum_j p_ja_j}{\sum_j p_jb_j}.\]
\end{lemma}

To see the lemma let $a=\sum_j p_ja_j$ and $b=\sum_j p_j b_j$ and
assume that for all $i$ we have $a_ib<b_ia$. Then also for all $i$ with $p_i> 0$ we
have $p_i a_i b< p_i b_i a$ and hence $b \sum_i p_i a_i<a\sum_ip_i
b_i$, a contradiction.

So there must be one $z$, such that $|z|/(1-\log \hat{g}^2_z )\ge
\bar{s}(g)/(1+H(\hat{g}^2))$. Using that $z$ in the bound of
theorem 6.1 yields the lower bound.
\qquad\end{proof}

The above bound decreases with the entropy of the squared Fourier
coefficients. This seems unnecessary, since the method of
theorem 4.1 suggests that functions with highly disordered Fourier coefficients
should be hard. This leads us to the next bound.

\begin{lemma}
 For all functions $f:\{0,1\}^n\times\{0,1\}^n\to\{-1,1\}:$
\[BQC(f)=\Omega\left(\frac{H_D(\hat{f}^2)}{\log n}\right),\]
where $H_D(\hat{f}^2)=-\sum_z \hat{f}_{z,z}^2\log\hat{f}_{z,z}^2$.
\end{lemma}

\begin{proof}
Consider any quantum protocol for $f$ with communication $c$.
As described in lemma 3.3, we can find a set of $2^{O(c\log
 n)}$ weighted rectangles so that their sum yields a function
$h(x,y)$ that approximates $f$ entrywise within error $1/n^2$.

Consequently, due to lemma 4.3, the sum of certain Fourier coefficients
of $h$ is bounded:
\[\log \sum_{z\in\{0,1\}^n}|\hat{h}_{z,z}|\le O(c\log n).\]
Also $-\sum_{z\in\{0,1\}^n}\hat{h}_{z,z}^2\log \hat{h}^2_{z,z}\le2
\log (1+\sum_{z\in\{0,1\}^n}|\hat{h}_{z,z}|)\le O(c\log n)$ due to lemma 2.12.

But on the other hand $||f-h||_2\le 1/n^2$, which we will use to
relate  $H_D(\hat{f}^2)$ to $H_D(\hat{h}^2)$.
We employ the following lemma.
\begin{lemma}
Let $f,h:\{0,1\}^n\times\{0,1\}^n\to\R$ with $||f||_2,||h||_2\le 1$. Then
\[\sum_{z\in\{0,1\}^n}|\hat{f}_{z,z}^2-\hat{h}_{z,z}^2|\le 3||f-h||_2.\]
\end{lemma}

Let us prove the lemma. Define
\[Min_z=\left\{\begin{array}{ll} \hat{f}_{z,z}& \mbox{ if }
   |\hat{f}_{z,z}|\le|\hat{h}_{z,z}|\\
 \hat{h}_{z,z}& \mbox{ if }
    |\hat{h}_{z,z}|<|\hat{f}_{z,z}|\end{array}\right.\]
and
\[Max_z=\left\{\begin{array}{ll} \hat{f}_{z,z}& \mbox{ if }
    |\hat{f}_{z,z}|>|\hat{h}_{z,z}|\\
 \hat{h}_{z,z}& \mbox{ if }
    |\hat{h}_{z,z}|\ge|\hat{f}_{z,z}|.\end{array}\right.\]
Then
$\sum_{z\in\{0,1\}^n}|\hat{f}_{z,z}^2-\hat{h}_{z,z}^2|=\sum_z
Max^2_z-Min^2_z$ and
\[||f-h||^2_2\ge
\sum_z (\hat{f}_{z,z}-\hat{h}_{z,z})^2=\sum_z(Min_z-Max_z)^2.\]
Due to the triangle inequality we have
\[\sqrt{\sum_zMin^2_z}+||f-h||_2\ge\sqrt{\sum_z Max_z^2}\]
and
\[\sqrt{\sum_zMin^2_z}\ge \sqrt{\sum_z Max_z^2}-||f-h||_2,\]
which implies
\[\sum_zMin^2_z\ge \sum_z
Max_z^2-2\sqrt{\sum_zMax_z^2}\cdot||f-g||_2,\]
and
\begin{eqnarray*}
  \sum_z Max^2_z-Min^2_z&\le&2\sqrt{\sum_zMax_z^2}\cdot||f-h||_2\\
&\le&2\sqrt{\sum_z\hat{f}_{z,z}^2+\hat{h}_{z,z}^2}\cdot||f-h||_2\\
&\le&2 \sqrt{2}||f-h||_2.\end{eqnarray*}
Lemma 6.7 is proved.

So the distribution given by the squared $z,z$-Fourier coefficients of $f$
is close to the vector of the squared $z,z$-Fourier coefficients of
$h$. Then also the entropies are quite
close, by the following fact (see theorem 16.3.2 in \cite{CT91}).
\begin{fact}
Let $p,q$ be distributions on $\{0,1\}^n$ with $d=\sum_z |p_z-q_z|\le
1/2$. Then $|H(p)-H(q)|\le d\cdot n-d\log d$.\end{fact}

Actually the fact also holds if $p,q$ are subdistributions, i.e.,
if they consist of nonnegative numbers summing up to at most 1.

So we get \[H_D(\hat{h}^2)\ge H_D(\hat{f}^2)-O(1/n).\]
Remembering that $H_D(\hat{h}^2)=O(c\log n)$ we get
\[H_D(\hat{f}^2)\le O(c\log n +1/n).\] This concludes the proof.
\qquad\end{proof}

If $f(x,y)=g(x\oplus y)$, then $H_D(\hat{f}^2)=H(\hat{f}^2)=H(\hat{g}^2)$.
Now we would like to get rid of the entropies in our lower bounds at
all, since the entropy of the squared Fourier coefficients is in
general hard to estimate. Therefore we would like to combine the
bounds of lemmas 6.4 and 6.6. The first holds for functions
$g(x\wedge y)$, the second for functions $g(x\oplus y)$.

\begin{definition}
A communication problem $f:\{0,1\}^n\times\{0,1\}^n\to\{-1,1\}$
can be reduced to another problem
$h:\{0,1\}^m\times\{0,1\}^m\to\{-1,1\}$,
if there are functions $a,b$ so that $f(x,y)=h(a(x),b(y))$ for all
$x,y$.\end{definition}

In this case the communication complexity of $h$ is at least as
large as the communication complexity of $f$.
Note that if $m$ is much larger than $n$, a lower bound which is a
function of $n$ translates into a lower bound which is a function of $m$,
and is thus ``smaller''. For more general types of reductions in
communication complexity see \cite{BFS86}.

If we can reduce $g(x\wedge y)$ and $g(x\oplus
y)$ to some $f$, then
combining the bounds of lemmas 6.4 and 6.6 gives a lower bound of
$\Omega(\bar{s}(g)/(1+H(\hat{g}^2))+H(\hat{g}^2)/\log n)$, which
yields corollary 1.2.

\subsection{A Bound Involving Singular Values}

We return to the technique of lemma 6.6. For many functions, like $IP_m$,
the entropy of the squared diagonal Fourier coefficients is small,
because these coefficients are all very small. We consider the entropy of a
vector of values that sum to something much smaller than 1 in cases.
Consequently it may be useful to consider other unitary
transformations instead of the Fourier transform.

It is well known that any quadratic matrix $M$ can be brought
into diagonal form by multiplying with unitary matrices, i.e., there are
unitary $U,V$ so that $M=UDV^*$ for some positive diagonal $D$. The
entries of $D$ are the singular values of $M$, they are unique
and equal to the eigenvalues of $\sqrt{MM^*}$, see \cite{B97}.

Consider a communication matrix for a function
$f:\{0,1\}^n\times\{0,1\}^n\to\{-1,1\}$. Then let $M_f$ denote the
communication matrix divided by $2^n$. Let
$\sigma_1(f),\ldots,\sigma_{2^n}(f)$ denote the singular values of
$M_f$ in some decreasing order. In case $M_f$ is symmetric these are just
the absolute values of its eigenvalues. Let $\sigma^2(f)$ denote the
vector of squared singular values of $M_f$. Note that the sum of the
squared singular values is 1.
The following theorem is a modification of lemma 6.6 and theorem 4.1.

\begin{theorem} Let $f:\{0,1\}^n\times\{0,1\}^n\to\{-1,1\}$
be a total Boolean function.

Then $BQC(f)=\Omega(H(\sigma^2(f))/\log n)$.

Let $\kappa_k=\sigma_1(f)+\cdots+\sigma_k(f)$.

If $\kappa_k\ge\Omega(\sqrt{k})$, then $BQC(f)=\Omega(\log (\kappa_k))$.

If $\kappa_k\le O(\sqrt{k})$, then
$BQC(f)=\Omega(\log (\kappa_k) / (\log(\sqrt{k})-\log(\kappa_k)+1))$.
\end{theorem}

\begin{proof}
We first consider the entropy bound and proceed similarly as in the
proof of lemma
6.6. Let $f$ be the considered function and let $h$ be the function
computed by a protocol decomposition with error $1/n^2$ consisting of
$P$ rectangles with $\log P=O(c\log n)$ for the communication
complexity $c$ of some protocol computing $f$ with error 1/3.

$M_f$ denotes the communication matrix of $f$ divided by $2^n$, let
$M_h$ be the corresponding matrix for $h$. Using the Frobenius norm on
the matrices we
have $||M_f-M_h||_F= ||f-h||_2\le 1/n^2$. Then also the singular values
of the matrices are close due to the Hoffmann-Wielandt theorem for
singular values, see corollary 7.3.8 in \cite{HJ85}.
\begin{fact} Let $A,B$ be two square matrices with singular values
  $\sigma_1\ge\cdots\ge \sigma_m$ and $\mu_1\ge\cdots\ge\mu_m$.
Then \[\sqrt{
\sum_i (\sigma_i-\mu_{i})^2 }\le ||A-B||_F.\]\end{fact}

As in lemma 6.6 we can use lemma 6.7 to show that the $L_1$-distance
between the vector of squared singular values of $M_f$ and the
corresponding vector for $M_h$
is bounded and fact 6.8 to show
that the entropies of the squared singular values of $M_f$ and $M_h$
are at most $o(1)$ apart.

It remains to show that $H(\sigma^2(h))$ is upper bounded by $\log
P$. Due to lemma 2.12 $H(\sigma^2(h))\le 2\log (1+\sum_i \sigma_i(h))$.
Due to the Cauchy Schwartz inequality we have
\begin{eqnarray*}
&&2\log(1+\sum_i\sigma_i(h))\\
&\le& 2 \log \sqrt{\sum_i\sigma^2_i(h)}\sqrt{rank(M_h)}+O(1)\\
&\le&\log rank(M_h)+O(1)\le\log P+O(1).\end{eqnarray*}

The last step holds since $M_h$ is the sum of $P$ rank 1 matrices. We
get the desired lower bound.

To prove the remaining part of the theorem we argue as in the proof of theorem 4.1 that
the sum of the selected singular values of $M_h$ is large compared to
the sum of the selected singular values of $M_f$, then upper bound the
former as above by the rank of $M_h$ and thus by $P$. The remaining
argument is as in the proof of theorem 4.1.
\qquad\end{proof}

Note that for $IP_n$ all
singular values are $1/2^{n/2}$, so the entropy of their squares is $n$, while the
entropy of the squared diagonal Fourier coefficients is close to 0,
since these are all $\langle IP_n,\chi_{z,z}\rangle^2=1/2^{2n}$. The log of the sum of all singular
values yields a linear lower bound. In this case the bounds of lemma 6.6 and
theorem 4.1 are very small, while theorem 6.10 gives large bounds.

Ambainis \cite{A01} has observed that theorem 6.10 can also be
deduced from a lower bound on the quantum communication complexity of
sampling \cite{ASTVW98}, using success amplification and an argument
relating the smallest number of singular values whose sum is at least
$1-\kappa_k^2/(4k)$ to the sum of the first $k$ singluar values in the
presence of sufficiently small error.

Note that theorem 6.10 does not necessarily generalize our other bounds in
the sense that the results obtained by using theorem 6.10 are better
for all functions.

We mention that the quantity $\sigma_1+\cdots+\sigma_k$ is known as the
Ky Fan $k$-norm of a matrix \cite{B97}. Well known examples of such
norms are the cases $k=1$, which is the spectral norm, and the case of
maximal $k$, known as the trace norm.
The Ky Fan norms are unitarily invariant for all $k$,
and there is a remarkable fact saying that if matrix $A$ has smaller
Ky Fan $k$-norm than $B$ for all $k$, then the same holds for \textit{any}
unitarily invariant norm. This leads to the interesting statement that
the Raz-type bound in theorem 6.10 for a function $g$ is smaller than the respective
bound for $f$ for all $k$, iff for all unitarily invariant matrix norms
$|||M_g|||\le |||M_f |||$. Under the same condition
the distribution $(\sigma_1^2(f),\ldots,\sigma_{2^n}^2(f))$ induced by the
singular values of $M_f$ majorizes the
distribution $(\sigma_1^2(g),\ldots,\sigma_{2^n}^2(g)) $ induced by $M_g$. This
implies that $H(\sigma^2(f))\le H(\sigma^2(g))$. Conversely,
considering the bounds in theorem 6.10: if the entropy bound for $g$ is
smaller than the entropy bound for $f$, then there is
a $k$, so that the Raz type bound for $k$ applied to $g$ is bigger
than the corresponding bound for $f$.

\subsection{Examples}

To conclude this section we give examples of lower bounds
 provable using the methods described by theorem 6.1 and corollary 1.2.
\begin{theorem}
$BQC(MAJ_n)=\Omega(n/\log n)$.
\end{theorem}

\begin{proof}
We change the range of $MAJ_n$ to $\{-1,+1\}$. Now consider the
Fourier coefficient with index $z=1^n$. $MAJ_n=g(x\wedge y)$ for a
function $g$ that is 1, if at least $n/2$ of its inputs are
one. W.l.o.g.~let $n/2$ be an odd integer. Thus any input to $g$ with
$n/2$ ones is accepted by both $g$ and $\chi_z$. Call the set of these
inputs $I$. Similarly every input
to $g$ with an odd number of ones larger than $n/2$ is accepted by
both $d$ and $\chi_z$ and every input
to $g$ with an even number of ones smaller than $n/2$ is rejected by
both $d$ and $\chi_z$. On all other inputs $g$ and $\chi_z$
disagree. Thus there are $|I|$ inputs more being classified correctly
by $\chi_z$ than those being classified wrong. The Fourier coefficient
$\hat{g}_z$
is $2 {n\choose n/2}/2^n=\Omega(1/\sqrt{n})$. So the method
of theorem 6.1 gives the claimed lower bound.\qquad\end{proof}

Note also that the average sensitivity of the function $g$ with
$MAJ_n(x,y)=g(x\wedge y)$ is $\Theta(\sqrt{n})$.

As another example we consider a function $g(x\wedge y\oplus z)$ with
a nonsymmetric $g$. Let $MED(a)$ be the middle bit of the median of
$n/(2\log n)$ numbers of $2\log n$ bits given in $a$. Let us compute a
lower bound on the average sensitivity of $MED$. For all inputs $a$
there are $\Theta(n/\log n)$ numbers bigger than the median and
smaller than the median each. For each number $p$ different from the
median we can switch a single bit to put the changed number below
resp.~above the median, shifting the median in the sorted sequence by
one position. For a random $a$ such a bit flip entails a change of the middle
bit of the median with constant probability. Hence the average
sensitivity of $MED$ is at least $\Omega(n/\log n)$. With corollary 1.2
this gives us a lower bound of $\Omega(\sqrt{n}/\log n)$ on the
bounded error quantum communication complexity of $MED(x\wedge y\oplus z)$.

\section{Application: Limits of Quantum Speedup}

Consider $COUNT^t_n(x,y)$.
These functions do admit some speedup by quantum protocols, this follows
from a black box algorithm given in \cite{BHT98} (see also
\cite{BBCMW98}), and the results of
\cite{BCW98} connecting the black box and the communication model.
\begin{lemma}
$BQC(COUNT^t_n)=O(\sqrt{nt}\log n )$.
\end{lemma}

Note that the classical bounded error communication complexity of all $COUNT_n^t$ is $\Theta(n)$, by a reduction from $DISJ_n$.
\begin{theorem} Let $t:\N\to\N$ be any monotone increasing function
with $t(n)\le n/2$. Then
\[BQC(COUNT^{t(n)}_n)\ge \Omega\left(\frac{t(n)}{\log t(n)}+\log
 n\right ).\]
\end{theorem}

\begin{proof}
  First consider  $COUNT^{n/2}_n$. This function is
equivalent to a function $g(x\wedge y)$, in which $g$ is 1 if the
number of ones in its input is $n/2$, and $-1$ else.
Consider the Fourier coefficient for $z=1^n$. For simplicity assume
that $n$ is even and
$n/2$ is odd. Then clearly
$\hat{g}_z=2 {n\choose n/2 }/2^n=\Omega(1/\sqrt{n})$. Thus the method
of theorem 6.1 gives us
the lower bound $\Omega(n/\log n)$. Note that finding this lower bound
is much easier than the computations in section 5 for $HAM_n^{n/2}$, since
we have to consider only one coefficient.

Now consider functions $COUNT^t_n$ for smaller $t$. The logarithmic
lower bound is obvious from the at most exponential speedup obtainable
by quantum protocols \cite{K95}.

Fixing $n/2-t$ pairs of inputs variables to ones and $n/2-t$ pairs of
input variables to zeroes leaves us with $2t$ pairs of free variables and
the function accepts if $COUNT^t_{2t}$ accepts on these inputs. Thus
the lower bound follows.
\qquad\end{proof}

Computing the bounds for $t=n^{1-\epsilon}$ yields corollary 1.3.

\section{Discrepancy and Weakly Unbounded Error}

The only general method for proving lower bounds on the quantum
bounded error communication complexity has been the discrepancy method
prior to this work. We now characterize the parameter $disc(f)$ in terms of the
communication complexity of $f$. Due to fact 2.8 we get for all
$\epsilon>0$
\begin{eqnarray*}
BQC_{1/2-\epsilon} (f)&=&\Omega(\log (\epsilon/disc(f)))\\
\Rightarrow BQC_{1/2-\epsilon}(f)-\log(\epsilon)&=&\Omega(\log
(1/disc(f))).\\
\mbox{Thus } PC(f)\ge QC(f)&=&\Omega(\log (1/disc(f))).
\end{eqnarray*}

\begin{theorem} For all $f:\{0,1\}^n\times \{0,1\}^n\to \{0,1\}:$

$PC(f)=O(\log(1/disc(f))+\log n)$.
\end{theorem}

\begin{proof}
Let $disc(f)=1/2^c$.
We first construct a protocol with public
randomness, constant communication, and error $1/2-1/2^{c+1}$, using
the Yao principle, and then switch to a
usual weakly unbounded protocol (with private randomness) with
communication $O(c+\log n)$ and the same error using a result of Newman.

We know that for all distributions $\mu$ there is a rectangle with
discrepancy at least $1/2^c$. Then the
weight of ones is $\alpha+1/2^{c+1}$ and the weight of zeroes is
$\alpha-1/2^{c+1}$ or vice versa on that
rectangle (for some $\alpha\in[0,1/2]$).
 
We take that rectangle and partition the rest of the communication
matrix into 2 more rectangles. Assign to each rectangle the label 0 or
1 depending on the majority of function values in that rectangle
according to $\mu$. The error of the rectangles is at most 1/2.
If a protocol outputs the label of the adjacent
rectangle for every input, the error according to $\mu$ is only $1/2-1/2^{c+1}$.

This holds for all $\mu$. Furthermore the rectangle partitions lead to
deterministic protocols with $O(1)$
communication and error $1/2-1/2^{c+1}$: Alice sends the names of
the rectangles that are consistent with her input. Bob then picks the
label of the only rectangle consistent with both inputs.

We now invoke the following lemma due to Yao (as in \cite{KN97}).

\begin{fact}
The following statements are equivalent for all $f$:

\begin{remunerate}
\item For each distribution $\mu$ there is a deterministic protocol for $f$
with error $\epsilon$ and communication $d$.

\item There is a randomized protocol in which both players can access a
public source of random bits, so that $f$ is computed with error
probability $\epsilon$ (over the random coins), and the communication
is $d$.
\end{remunerate}
\end{fact}

So we get an $O(1)$ communication randomized protocol with error
probability $1/2-1/2^{O(c)}$ using public randomness.
We employ the following result from \cite{N91} to get a protocol with private
randomness.

\begin{fact}
Let $f$ be computable by a probabilistic protocol with error
$\epsilon$, that uses public randomness and $d$ bits of
communication. Then
$BPC_{(1+\delta)\epsilon}(f)=O(d+\log(\frac{n}{\epsilon\delta}))$.
\end{fact}

We may now choose $\delta=1/2^{O(c)}$ small enough to get a weakly
unbounded error protocol for $f$ with cost $O(c+\log n)$.\qquad\end{proof}

Let us also consider the quantum version of weakly unbounded error
protocols.

\begin{theorem} For all $f$: $PC(f)=\Theta(QC(f))$.
\end{theorem}

\begin{proof}
The lower bound is trivial, since the quantum protocol can simulate
the classical protocol.

For the upper bound we have to construct a classical protocol from a
quantum protocol. Consider a quantum protocol with error
$1/2-\epsilon\le 1/2-1/2^c$ and communication $c$. Due to lemma 3.4 this
gives us a set of $2^{O(c)}$ weighted rectangles, such that the sum of the
rectangles approximates the communication matrix entrywise within
error $1/2-\epsilon/2$. The weights are real $\pm \alpha$ with absolute
value smaller than 1. Label the $-\alpha$ weighted rectangles with 0
and the other rectangles with 1, and add $(1/2)/\alpha$ rectangles
covering all inputs and bearing label 0. This 
clearly yields a majority cover of size
$2^{O(c)}$, which
is equivalent to a classical weakly unbounded error protocol using
communication $O(c)$ due to fact 2.3.
\qquad\end{proof}

It is easy to see that there are weakly unbounded error protocols for
$HAM_n^{t}$, $MAJ_n$, and $COUNT^t_n$ with cost $O(\log n)$. $MAJ_n$
is even a complete problem for the class of problems computable with polylogarithmic
cost by weakly unbounded error protocols.
\begin{lemma} For $f\in\{MAJ_n,HAM_n^{t},COUNT_n^t\}:$
$\max_\mu\log( 1/disc_\mu(f) )=O(\log n)$.
\end{lemma}

\section{Discussion}

In this paper we have investigated the problem of proving lower bounds
on the bounded error quantum communication complexity. As opposed to
previous approaches our methods are both general and make use of
the quantum properties of the protocols (i.e., do not follow the
scheme of simulating a bounded error quantum protocol by an unbounded error classical protocol and
employing a lower bound method for the latter). Our results are strong
enough to show separations between unbounded error classical
and bounded error quantum communication resp.~between
quantum nondeterministic and quantum bounded error communication.

Our results do not address the more powerful model of quantum
communication complexity with prior entanglement \cite{CB97, CDNT98}. It
would be interesting to obtain similar results for this model. Recently
an improved lower bound (compared to \cite{CDNT98}) for the complexity
of $IP_n$ in this model has been obtained in \cite{NS02}. These bounds
do not show hardness under a distribution like in the second statement
of fact 2.8, though. So constructions similar to that of theorem 6.1
remain unknown for the model with prior entanglement.
 
More recently Razborov \cite{R02} has obtained much stronger lower bounds
on the quantum communication complexity of $g(x\wedge y)$
for {\it symmetric} functions $g$, almost tightly characterizing the quantum
bounded error communication complexity of these functions, even in the
model with prior entanglement. This give a $\Omega(\sqrt{n})$ lower
bound for $DISJ_n$, previously superlogarithmic bounds for this
function were known only for the cases when
strong restrictions on the interaction are imposed \cite{KNTZ01}
or when the error probability is extremely small \cite{BW01}.
Razborov's techniques are based on showing good lower
bounds on the minimal trace norm (sum of singular values) of matrices
approximating the communication matrix, similar to the approach in
theorem 6.10. These new results can be used to show that in our
corollary 1.3 actually the upper
bounds for $COUNT^t_n$ are tight.

The lower bound methods of this paper can also be applied to other
types of functions, see sections 5 and 6.4.
It would be interesting to find tighter lower bounds
for these functions and to extend our results to the model with prior entanglement.

Finally we would like to know if quantum bounded error communication
can ever be more than quadratically smaller than classical bounded
error communication for total functions. A first step to resolve this
problem would be to show a lower bound in terms of (one-sided) block
sensitivity on the quantum bounded error complexity of all functions
$g(x\wedge y)$ (with nonsymmetric $g$).

\section*{Acknowledgements}

The author wishes to thank Ronald de Wolf for bringing \cite{R95}
to his attention and for lots of valuable discussions, and Andris Ambainis for pointing
out a mistake in an earlier version of the paper.

\end{document}